# A Framework for Techniques for Information Technology Enabled Innovation

**Yajur Chadha**
Research School of Management
The Australian National University
Canberra, Australia
Email: u5479605@anu.edu.au

**Aditi Mehra**
Research School of Management
The Australian National University
Canberra, Australia
Email: u5485521@anu.edu.au

**Shirley Gregor**
Research School of Management
The Australian National University
Canberra, Australia
Email: shirley.gregor@anu.edu.au

**Alex Richardson**
Research School of Management
The Australian National University
Canberra, Australia
Email: alex.richardson@anu.edu.au

## Abstract

Australia is seen as lagging in the innovation that is needed for corporate success and national productivity gains. There is an apparent lack of consistent and integrated advice to managers on how to undertake innovation. Thus, this study aims to develop and investigate a framework that relates innovation practices to the type of innovation outcome, in the context of Information Technology (IT) enabled innovations. An Innovation Practice Framework was developed based on the Knowledge-Innovation Matrix (KIM) proposed by Gregor and Hevner (2015). Eleven commonly used innovation techniques (practices) were identified and placed in one or more of the quadrants: invention, advancement, exaptation and exploitation. Interviews were conducted with key informants in nine organisations in the Australian Capital Territory. Results showed that the least used techniques were skunk works and crowdsourcing. The most used techniques were traditional market research, brainstorming and design thinking. The Innovation Practice Framework was given some support, with genius grants being related to invention outcomes, design thinking with exaptation, traditional R&D with advancement and managerial scanning with exploitation. The study contributes theoretically with the new Innovation Practice Framework and has the potential to be useful to managers in showing how benefits can be gained from a range of innovation practices. Further work is in progress.

## 1  Introduction

Innovation is seen as a key driver of corporate success (Cardozo et al. 1993). Innovation in organisations can boosts productivity, create and sustain a competitive advantage and promote economic growth at the national level (Danneels 2002). However, in Australia, the competitive advantage of Australian organisations is seen to be diminished by relatively greater adoption of existing market innovations compared with new-to-market innovations (Department of Industry 2014). The Australian Innovation System Report 2014 found that "Australia's innovation system is a mid-range performer among Organisation for Economic Co-operation and Development (OECD) countries. The evidence suggests that our innovation performance is lagging, potentially leaving us less resilient to future global shocks".

Managers in Australian organisations, as elsewhere, may be inhibited by the lack of consistent advice on how to go about innovating, with a bewildering array of terminology leading to confusion. For example, innovation stakeholders are advised to be ambidextrous: to engage in simultaneous exploration and exploitation innovation strategies. More advice addresses radical-incremental, discontinuous-incremental, or breakthrough-incremental distinctions. Recommended innovation practices include skunk works, experimentation, prototyping, design science, design thinking, open innovation, crowdsourcing, innovation ecosystems – the list is almost endless. Confusion is heightened by the speed and reach of change associated with the current phenomena of digital disruption (Weill and Woerner 2015) and the limited knowledge on how innovations occur in this context.

A limited amount of work has addressed the question of whether different innovation techniques are associated with different types of innovation. Koen et al. (2014) showed that the processes for radical innovations differed from those for incremental innovations. Our study aims at adding to the sparse literature on this topic. The aim of our study is to develop and investigate a framework that relates innovation practices to the type of innovation outcome, in the context of Information Technology (IT) enabled innovations.

By innovation practice (or technique) we mean the tools and techniques that are used in carrying out innovation. The unit of analysis in the current study is an innovation project.

The study has both theoretical and practical significance. Theoretically it adds to the literature on processes for IT-innovation in the age of digital disruption. Practically, the results can inform managers on how different innovation techniques can be used effectively for different types of innovation outcome.

## 2 Literature Review

### 2.1 Introduction

The literature on innovation is huge and a review is beyond the scope of this paper. Recent work that provides an overview of IT related work includes Chua et al. (2013), Luo et al. (2012), Yoo et al. (2012). Luo et al. (2012) use absorptive capacity (ACAP) as a lens, defined by two dimensions routines and knowledge base, to understand how the knowledge of a software organisation influences its radical IT innovations during a technological breakthrough. Chua et al. (2013) study the radical innovations (RI) that rely on digital technologies and propose the technology, application and market trend model (TAMT) of RI that emerges from interactions between these three aspects. On the other hand, Yoo et al. (2012) discuss how pervasive digital technology, while rapidly being adopted by organisations, reshapes them. They discuss three traits of innovation – the importance of digital technology platforms, emergence of distributed innovations, and the prevalence of combinatorial innovations and the significant changes they make to organisational practices.

### 2.2 Knowledge Innovation Matrix (KIM)

The KIM model (Figure 1) has been chosen for this study primarily because it distinguishes between processes and activities that occur with different types of innovations based on a strong analytic framework. Innovations can be classified in a number of ways and this matrix provides a more detailed view than the dichotomous classifications of exploration and exploitation suggested by March (1991). The KIM model arose from the design science innovation framework in Gregor and Hevner (2013) and was developed initially for the processes at the front-end of innovation, where opportunities are identified, ideas generated, and concepts or prototypes developed for further stages of development (Koen et al. 2014a, 2014b).

The matrix has two dimensions, namely the *knowledge (solution) maturity* dimension and the *application domain (problem) maturity* dimension. The knowledge maturity dimension recognizes the importance of ideas, new insights, new knowledge, technological know-how, new knowledge and learning, whereas the application dimension recognizes opportunities, tasks and problems, markets, needs and fields. The matrix has four quadrants - *invention, advancement, exaptation* and *exploitation.* Gregor and Hevner (2015) show how different patterns of innovation practice are expected to arise in each quadrant based on theories of innovation, creativity and technology adoption (Amabile 1996; Andriani, Carignani and Kaminska-Labbe 2013; Cropley and Cropley 2010; Csikszentmihalyi 1997; Rogers 2005; von Hippel and von Krogh 2013).

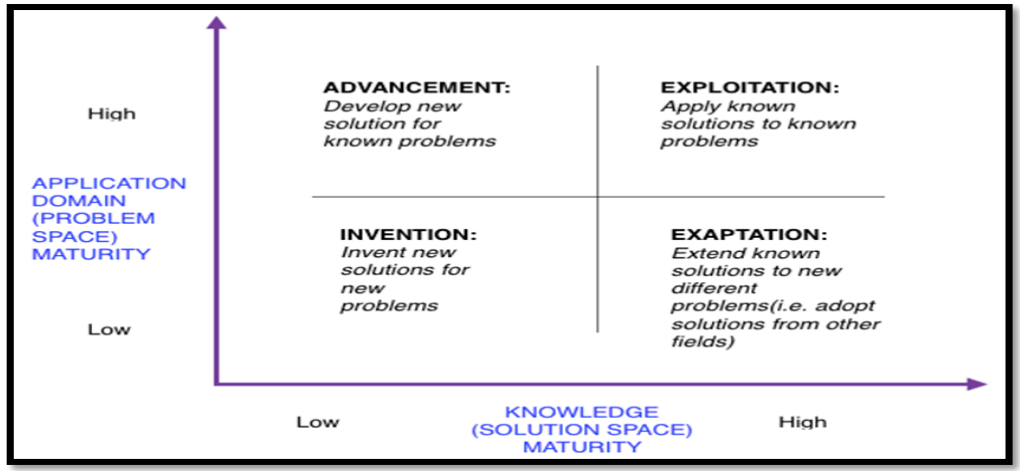

Figure 1: Knowledge-Innovation Matrix (Gregor and Hevner 2014 2015)

The invention quadrant comprises those innovations that are "new-to-the world", where both the idea of the problem and the knowledge required for its implementation have not been identified before. The advancement quadrant includes innovations where an existing problem is enhanced by developing a superior solution. The exaptation quadrant comprises those innovations where the knowledge used to implement one application or need is used for another need or problem in a completely different context. Existing knowledge or solutions are used for a completely different purpose. The exploitation quadrant includes those innovations that are adopted from existing innovations. Known solutions are applied to known problems with minor customisation – the innovation may be "new-to-us" rather than "new-to-the world".

In terms of the March (1991) distinction between exploitation and exploration, the exploitation quadrant is similar to March's concept of exploitation, while the other three quadrats are closer to exploration.

### 2.3 Prior Work on Innovation Practices

March (1991) suggests that an organisation should adopt a mixture of exploration and exploitation practices/techniques in order to survive and maintain prosperity. It has been shown that firms that behave ambidextrously, balancing exploration and exploitation well are nine times more likely to achieve breakthrough products and processes than others, even while sustaining their existing businesses (O'Reilly and Tushman 2004).

Carvalho and dos Reis (2012) list 67 techniques that contribute to idea generation and ultimately creation of innovation. A survey by PricewaterhouseCoopers (PWC 2015) also shows that organisations use multiple tools and techniques for innovation. The results of this survey indicate that most organizations use traditional idea generation techniques such as direct customer observation, traditional market research, feedback from sales and customer support, idea work-out sessions, technology road mapping and other sources (see Figure 2).

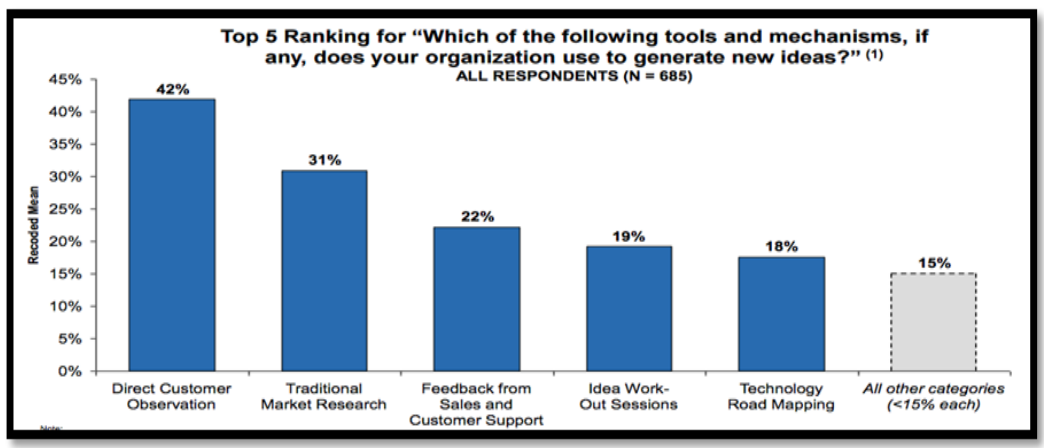

*Figure 2: Tools and Mechanisms (adapted from PWC 2015)*

Few prior studies have differentiated practices by type of innovation. An exception is the study by Koen et al. (2014) that focused on front-end innovation and examined activities in terms of the New Concept Development model (NCD) to show that the processes for radical innovations differ from those for incremental innovations.

# 3  Research Method

This study is a part of larger action design research project (Sein et al. 2011). In this step of the study, we develop a framework for literature and investigate its applicability through interviews.

As a starting point for the study we use the Knowledge Innovation Matrix (KIM) framework for innovation types proposed by Gregor and Hevner (2014; 2015) for the front end of innovation (Koen et al. 2014). We extend their work by more precisely identifying the innovation techniques that are expected to be most applicable for the different types of innovation in the four quadrants in KIM: invention, advancement, exaptation, and exploitation. The KIM-techniques framework that results is used as a base for interviews with stakeholders in organisations in the Australian Capital Territory (ACT). At this point in our work-in-progress study only a small number of interviews have been conducted but some interesting results have been observed. In later phases of our work, we expect to further analyse data, carry out case studies and refine the KIM-techniques framework as necessary.

# 4  Innovation Practice Framework

This section describes the techniques that we chose for investigation and our new innovation practice framework. Techniques were chosen that are in common use and range over a number of innovation categories (see Figure 3).

## 4.1  Selection of Techniques

There are countless methods/techniques that are used for the implementation of any kind of innovation. With the help of a literature review, eight different techniques widely used for innovation were identified and defined. These eight techniques are *lead user method*, *skunk works*, *genius grants*, *design thinking*, *benchmarking*, *managerial scanning*, *crowdsourcing* and *traditional research and development (R&D)*. These eight techniques were selected because they are frequently used and have led to successful outcomes in organisations. This list was compared with the techniques identified in the study by PWC (2015) and expanded to include *traditional market research*, *brainstorming* and *technology road mapping*. This final selection of 11 techniques aligns with emerging technology trends and the growing need for organisations to innovate.

On the basis of their outcomes and implementation process and the theories described in Gregor and Hevner (2015), the 11 techniques were classified into the four quadrants (Figure 3) of the Knowledge Innovation Matrix (Figure 1). The placement of the techniques resulted from discussion amongst the authors until consensus was reached. It was realized that techniques could be used in more than one quadrant at different points in a project. The technique of brainstorming was placed across all the four quadrants, as its use is so ubiquitous.

## 4.2  Quadrant Analysis

We begin by defining each of the 11 innovation techniques, explaining the reason for its placement within the KIM (as shown in Figure 3) and giving a related example of IT innovation using each technique. Those techniques applying primarily to one quadrant only are discussed first.

### 4.2.1  Advancement Quadrant

**Traditional R&D**

Traditional research and development is two-step process in which 'research' is a systematic approach of gaining new knowledge or building better understanding of concepts and 'development' is the process of applying and using the knowledge gained through research to generate new ideas that can be applied to the organisation.

This technique is placed in the advancement quadrant because it helps in developing new technology for improving existing products. Scientists have good knowledge of the application domain but the technology required for addressing problems needs to be improved.

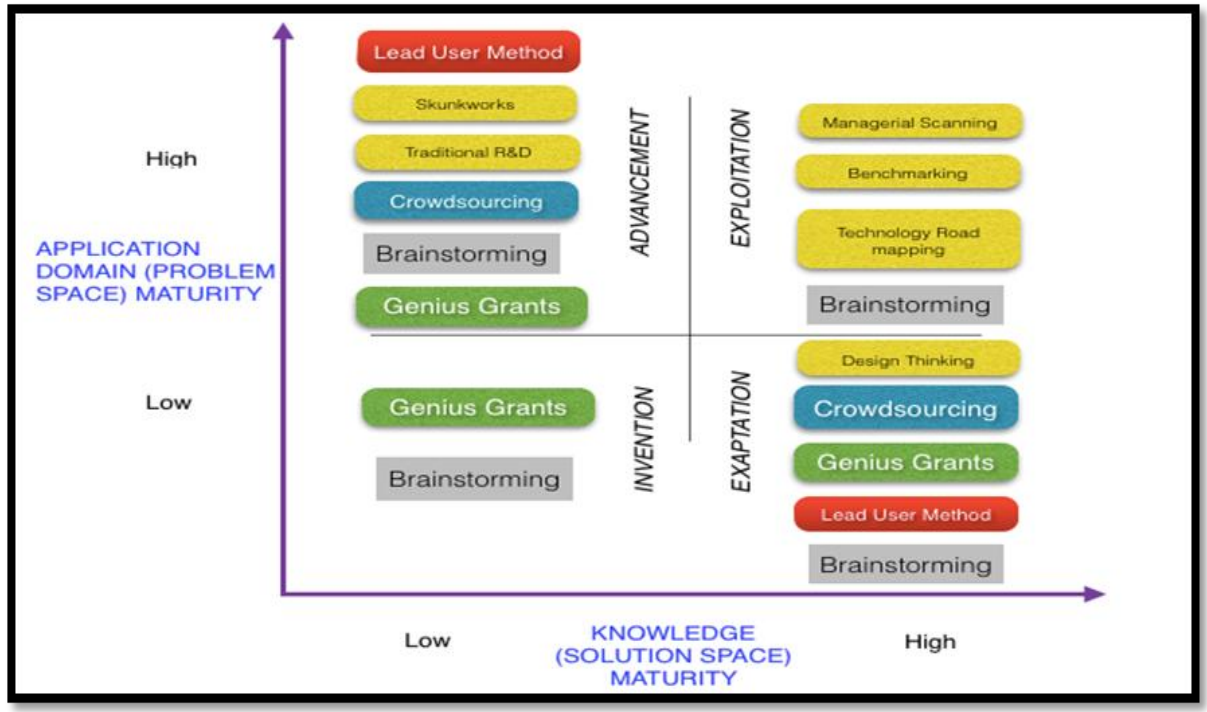

*Figure 3: Innovation Practice Framework with Illustrative Techniques*

- Example: Apple

One successful example is the iPad, where scientists used traditional R&D to develop new technology for the launch of different versions of the iPad with new and improved features such as reduced size, improved processor, reduced display and improved operating system (Apple 2015).

**Skunk works**

Skunk works is a technique where a small group of scientists, engineers or other personnel are isolated by the top management from the influence of the rest of the organisation, to tackle specific problems and commercialise solutions (Fosuri and Ronde 2009; Gwynne 1997).

Skunkworks has been positioned in the advancement quadrant as it involves the use of scientists who are equipped with high-end research material and are aware of the application domain but the technology needed to implement the application domain needs to be investigated.

- Example: Lockheed Martin

A famous skunk works project is the development of **Lockheed Martin's Mirror World** that modifies video game technology to create a highly accurate 3-D virtual world, combining both visual and geospatial elements (Mirror World 2015). This example is considered advancement, as these innovations modified the existing technology and involved the use of qualified scientists.

**Traditional Market Research**

Traditional market research involves the collection of customer needs data through surveys and customer feedback sessions. With the help of the collected data, the internal staff and research team make improvements in existing products to satisfy customer needs (Churchill et al. 2009). This technique rarely leads to creation of new products or services.

This technique has been placed under the advancement quadrant because it helps in improving the existing products and services in response to customer feedback.

- Example: Google, Apple and so on

Many organisations use online platforms and other market research approaches to further improve their existing products. For example, an airline company uses feedback from frequent air travellers with the experience of mobile applications to improve their own mobile application thereby, improve frequent flyer customer satisfaction.

### 4.2.2 Exaptation Quadrant

**Design Thinking**

Design thinking is considered a combination of analytical and intuitive thinking that extends beyond the look and feel of the product and is comprised of tools and frameworks that reflect its concern with human experience. It involves customer-understanding, visualisations, prototyping and strategy design (Euchner 2012; Gobble 2014).

Design Thinking is placed in the exaptation quadrant as it allows integration of the customer needs with the creative ideas of the employees. Thus, the application domain maturity is low and knowledge or technology to implement is high.

- Example: IDEO and Bank of America

An exaptation of IDEO and Bank of America is a savings account named "**Keep the change**". IDEO along with a team at the bank identified the customer behaviour of putting change coins in a jar after paying cash, and later taking the jar and depositing those coins in a savings account. Thus, Bank of America decided to implement this behaviour by modifying its IT systems to incorporate this functionality into debit card accounts. Whenever a customer uses a debit card to make a purchase the total is rounded up to the nearest dollar and the difference is deposited in their savings account. This invention is considered an exaptation as the customer behaviour of using a jar to store the change coins was used for a new product in the banking industry (Brown 2008).

### 4.2.3 Exploitation Quadrant

**Benchmarking**

Benchmarking involves measuring and comparing an organisation's operations, practices and performance against other organisations. This is a market-based management tool to identify the best practices that produced superior results elsewhere and then replicate these practices to improve one's own competitive advantage. (Copp 2002; Sekhar 2010; Vorhies and Morgan 2005).

- Example: Taobao

A successful example of benchmarking is the Chinese e-commerce giant Alibaba creating Taobao as a defensive move against the US rival eBay who started operations in China. According to Greeven et al. (2012), Taobao bested eBay in China and now holds 80% of China's e-commerce market. Taobao benchmarked the concept of online auctions from eBay and made customisations to EBay's practices to suit Chinese customers.

**Managerial Scanning**

This technique is used in the early stages of strategic planning process as a means to identify the important strategic questions facing the organisations, where the organisation is viewed as a member of a much larger network and related to other institutions in the network. It identifies the strategic profiles of the member organisations in the network so that future courses of action can be taken. Unlike benchmarking, managerial scanning does not measure operational success or evaluate competitive abilities. Its main focus lies external to the home organisation with little internal reference (Cancellier et al. 2014).

As this technique involves adopting the practices and using them in their own organisation and where every person contributes in the innovation, it has been placed in the exploitation quadrant.

- Example: Semco

One famous example of managerial scanning is the adoption of a fluidic structure by Semco in order to save itself from bankruptcy. The CEO surveyed the strategic profiles of other organisations and found that in order to keep Semco abreast with the technology, innovation needs to be spread across the whole organisation rather than a single department. The adoption of a fluidic structure enabled people to participate in any project of their choice and contribute to innovation. Thereafter, every employee at Semco participates in every organisational decision (Sawyer 2007).

**Technology Road mapping**

Technology road mapping is a technique that is used within an industry to support strategic long term planning and enables the organisations to align investments in technology and the new development of capabilities, so that they are able to maximise their ROI (Bernal et al. 2009; Phaal et al. 2004).

Technology road mapping essentially uses a process to identify the time point at which competitors will probably launch innovations that are similar to the one's planned by them. Thus, this technique has been placed in the exploitation quadrant.

- Example: Motorola

Motorola first developed technology road maps for aligning the development of their products and their supporting technologies. This technique provided Motorola a means of communicating to the design and development engineers and to the market personnel, which technologies will be requiring development and application for future products. It was introduced to maintain a balance between short and long-range issues, strategic and operational matters with technology (Willyard and McClees 1987).

### 4.2.4 Multiple Quadrants

Some of the techniques can be placed across multiple quadrants:

**All four Quadrants**

**Brainstorming**

Brainstorming is a problem solving technique that involves contribution of ideas from all the members in a group and building over the ideas contributed by other individuals in an attempt to devise a solution for a problem (Paulus and Yang 2000). This technique can be used as a part of any kind of innovation and thus, has been placed across all the four quadrants.

**Advancement and Exaptation Quadrants**

**Crowdsourcing**

This is an emerging Web 2.0-based phenomenon that helps to accomplish a task by opening up its completion to broader sections of the public through different online platforms (Ranard et al. 2014; Zhao and Zhu 2014).

Crowd sourcing can be placed in the advancement quadrant because the application to be developed or improved is known. Thus, the application domain is high. However, the knowledge to implement that application is not well known. Thus, the knowledge maturity is quite low. As the knowledge maturity is low, the idea is made open to the general public.

- Example: Wikipedia

Wikipedia.org, whose users have access to a quantitatively and qualitatively growing pool of knowledge, is a successful crowd sourcing venture. By disclosing the intangible resource "knowledge", the crowd creates value for the general public. During its development, scientists were aware of the fact that the goal of this venture was to create a public document repository (Hammon and Hippner 2012).

This technique can also be placed in the exaptation quadrant as the ideas used in one industry can prove to be useful for another purpose in another industry.

- Example: InnoCentive

Werner Mueller, a chemist, opened a lab after retirement because of his interest in the field of chemistry. Mueller helped an organisation to find a new use for a compound and now uses his lab for answering questions on website InnoCentive. Thus, the lab that was created as a source of interest was re-used for crowdsourcing solutions to those in need (Tapscott and Williams 2006).

**Lead User Method**

Henkel and Jung (2010) refer this as the 'technology-push lead user method', where the inventor (technology owner) leapfrogs the manufacturer and turns directly to potential users. In simple words, the inventor addresses users who are ahead of important market trends and benefit highly from products satisfying those needs – lead users. Thus, the lead user method goes beyond the customer-centred approaches, seeking inputs not only from customers but lead users, who have advanced needs that preview the future needs of general market place (Eisenberg 2011).

As per the definition, the lead user method can be used for developing new products by re-using the existing knowledge being used in one industry for a completely different purpose. Thus, it has been placed under the exaptation quadrant.

- Example: Texas Instruments

One of the successful lead user projects is that by Texas Instruments. They used the digital mirror device, originally used for printing airline tickets, to decrease the size and cost of digital video projectors after observing the issues faced by customers when using bulky projectors. They used the available technology of digital mirror device for a completely new purpose (Hornbeck 1993).

This technique can also be place in the advancement quadrant because it can lead to improvements in existing products or solutions in response to market needs when the technology to implement the change is low.

- Example: Sun Microsystems

Sun Microsystems recognised the market need for a smart appliance platform requiring portable code for network and distributed computing, which could not be fulfilled using C++, as the language did not support that functionality. On missing out on PC consumer market, Sun Microsystems (now part of Oracle) developed an object oriented programming language Oak, now known as Java, that can be used around different platforms and helped in capturing the PC consumer market. It is placed in the exaptation quadrant because developers used already available object oriented programming concepts for development of Java, which is an enhanced and evolved version of C++ and C (Bottoms 1995).

## Invention, Advancement and Exaptation Quadrants

### Genius Grants/Underground Innovation

Genius grants means that time is granted to employees by the organisation to work on individual innovation projects apart from their normal routine work. This practice is openly allowed by the organisation. A similar practice is bootlegging, which is a clandestine bottom-up activity hidden from the top management of the organisations (Criscuolo et al. 2014; Masoudnia and Szwejczewski 2012).

This particular technique is placed in the invention quadrant because genius grants are used when the person has no specific knowledge or technology to implement the idea but still comes up with a completely new invention.

- Example: 3M

Post-it notes developed by 3M is one such invention. Instead of a strong adhesive to be used in the aerospace industry, 3M scientists inadvertently developed a light adhesive that left no residue. Since the light adhesive had no formal use, scientists at 3M bootlegged to develop a product to put it to use that lead to the development of Post-it notes. In simple words, there was low application domain knowledge and low knowledge for putting the light adhesive to use. Thus, it is considered an invention (Owens 2011).

This particular technique can also be placed in the advancement quadrant as it involves scientists who are aware of the goal of innovation but do not have the required technology for carrying out innovation.

- Example: Google

Google's Gmail is a successful outcome of Google's 20 per cent time policy. After its launch, Gmail was the first email to contain a 1 GB storage space and a search option. It was an improvement over existing Yahoo Mail and Microsoft Hotmail web-based email. It is considered an advancement because the scientist was aware of the application domain but did not yet have the technology to develop an improved web-based email (McCracken 2014).

Genius grants can also be placed in the exaptation quadrant when solutions are already being used for one purpose in one industry and can be re-used for a new purpose in a different industry.

- Example: Pacific Tech

NuCalc by Pacific Tech is another example of IT innovation. It is also known as the Graphing Calculator 1.0 that was an improved version of Milo and Frame Maker, which were Apple products. The build for this software originally started as an Apple project but was later scrapped by the management and was pursued clandestinely by Pacific Tech. Since this project was based on multiple apple projects, it is considered an exaptation (Avitzur 2004).

## 5   Framework Investigation

The aim of this exploratory study was to investigate whether different types of practice would be associated with different types of innovation outcomes, as suggested by the KIM framework. Due to the nature of the research, face-to-face interviews using a questionnaire consisting of open and closed questions were considered to be appropriate method of research (Lee R. 1993; Lee R. and Renzetti C. 1990). Interviewees were enlisted in the study with the aid of the Canberra Innovation Network (CBRIN) and snowball sampling, whereby interviewees referred the interviewer to other people who matched the criteria. There was one interviewee per organisation. A semi-structured interview protocol was used and interviewees were asked about each technique in turn. For each technique, the technique was briefly explained, then interviewees were asked where and how they had used the technique, and for associated benefits, issues and barriers. Data was analysed using open coding techniques.

## 6   Results to Date

This section highlights the initial results from nine interviews. The nine organisations interviewed were a mix of start-up, government and large private organisations. The interviewees ranged from young entrepreneurs to senior level employees.

The initial results indicate that ACT organisations consider innovation important for their organisation but many lack a formal process to carry out innovation activities. Moreover, the interviews indicated that most organisations use more than one technique for the completion of a single innovation project.

Figure 4 shows the relative usage of each technique. The least used techniques were skunk works and crowdsourcing. Organisations reported that their organisation structure was such that they lacked the need to use skunk works whereas crowdsourcing was associated with a low rate of success. The most used techniques were traditional market research and brainstorming. Design thinking is the third most used technique and it is interesting that it has been taken up so well, perhaps due to recent popularization of the concept. What is a little surprising is that only two-thirds or fewer of the organisations use technology road-mapping, benchmarking or managerial scanning of competitors' behaviour, when these techniques would be expected in all organisations that expect to remain competitive.

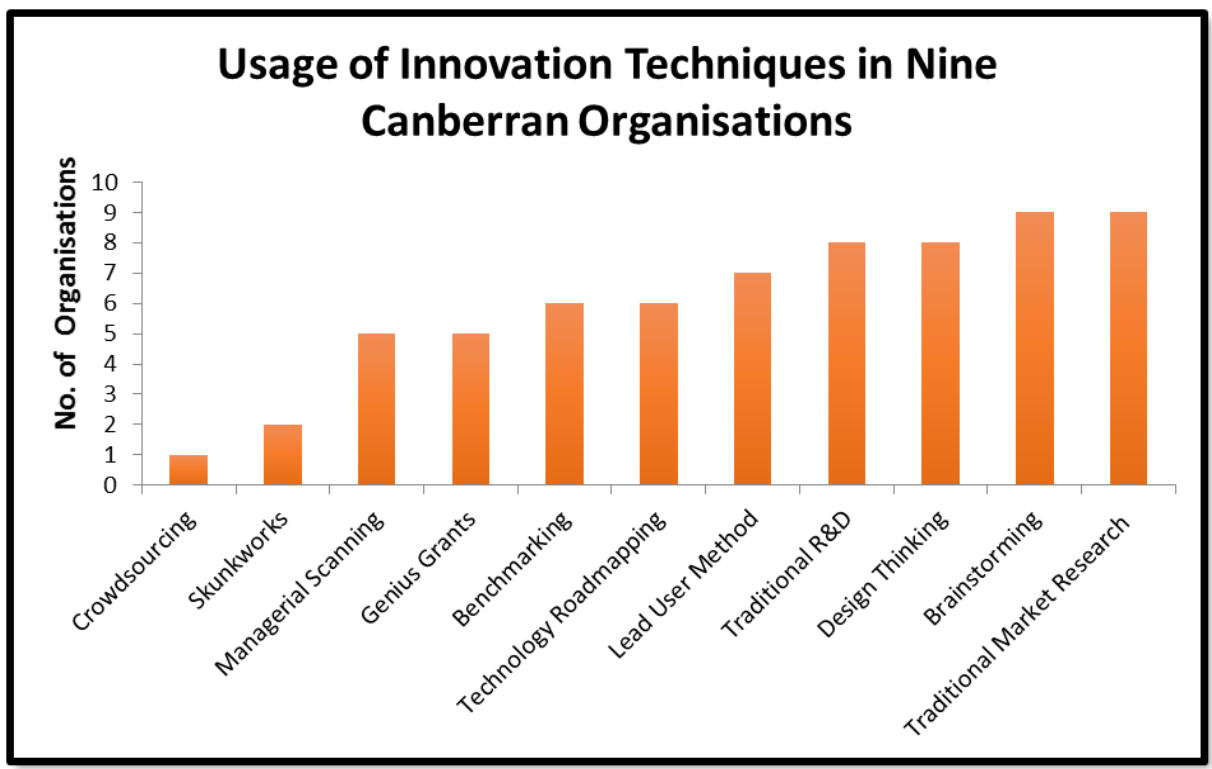

*Figure 4: Usage of Techniques*

Open coding of the qualitative data obtained in response to a question asking for the benefits realised from each technique allowed a check on whether techniques had been placed in the appropriate quadrant in the KIM Innovation Practice Framework (Figure 3). Note that not all answers contained terms that allowed a placement in any quadrant. The open-ended nature of the interview questions meant the interviewees spoke to some degree of extraneous issues.

The placement of benchmarking and managerial scanning techniques in the exploitation quadrant of the KIM was supported. Two organisations reported that the use of the benchmarking technique helped them in tracking their peers, identifying gaps and adapting the practices to use within their own organisation. Two interviewees reported that the use of managerial scanning technique helped them to identify the best practices and strengths of other organisations so that they can be replicated in their own organisation.

The placement of the traditional R&D technique in the advancement quadrant was supported. Two organisations reported that the traditional R&D technique is renewable, helps in upfront identification of customer needs and gives a whole new perspective in making improvements to business.

The placement of the genius grant technique in the innovation quadrant was supported to some extent. One organisation reported that using the genius grants technique in the organisation has helped them to tap into the enthusiasm of the employees to explore beyond current boundaries. This has helped them to generate new ideas that have resulted in interesting innovations.

The placement of design thinking in the exaptation quadrant was supported. Three organisations reported that the use of design thinking within their organisation has helped them to gain new business insights that are more customer-focussed and to develop a self-selling product (as it takes into account what customer wants). This helps them in gaining understanding about how new products need to be adapted according to customer needs.

Barriers reported by organisations while using the benchmarking technique were the difficulty of making comparisons with like organisations and having to focus on the measurement of different dimensions. Organisations reported that tracing a developed technology back to business requirements and the costs incurred were barriers while using Traditional R&D. Keeping people on track, restraining people from giving personal opinions rather than facts and avoiding conflicts were some of the barriers reported while using brainstorming.

# 7　Discussion and Future Directions

The aim of our exploratory study was to develop and investigate a framework that relates innovation practices to the type of innovation outcome, in the context of Information Technology (IT) enabled innovations. An Innovation Practice Framework was developed based on the Knowledge-Innovation Matrix (KIM) proposed by Gregor and Hevner (2015). Eleven commonly used innovation techniques were identified and placed in one or more of the quadrants: invention, advancement, exaptation and exploitation. The techniques were the lead user method, skunk works, genius grants, design thinking, benchmarking, managerial scanning, crowdsourcing, traditional research and development (R&D), traditional market research, brainstorming and technology road mapping.

Semi-structured interviews were conducted with key informants in organisations in the ACT. Results showed that the least used techniques were skunk works and crowdsourcing. The most used techniques were traditional market research, brainstorming and design thinking. The Innovation Practice Framework was given some support, with genius grants being related to invention outcomes, design thinking with exaptation, traditional R&D with advancement and managerial scanning with exploitation.

The study contributes with the Innovation Practice Framework, which gives a finer-grained well-based theoretical view of how innovation practices can be related to different innovation outcomes compared with prior dichotomous views such as that of exploration-exploitation or radical-incremental. The work has the potential to be useful to managers in showing how benefits can be gained from a range of innovation practices.

The study is work-in-progress and more interviews are being conducted to refine the results. Further analysis of data and case studies are planned.

## Acknowledgements

The authors would like to express their gratitude to Canberra Innovation Network (CBRIN) for their support in finding different organisations to interview.

## Copyright